\documentclass[twocolumn,prl,tighten,floatfix,showpacs]{revtex4}
\usepackage{graphicx}

\def\<{\langle}
\def\>{\rangle}

\def\be{\begin{equation}}
\def\ee{\end{equation}}

\begin{document}
\preprint{cond-mat} \title{Projective approach to the entanglement entropy of 1-$d$ fermions}

\author{G. C. Levine* and B. A. Friedman$^\dagger$}

\address{*Department of Physics and Astronomy, Hofstra University,
Hempstead, NY 11549}
\address{$^\dagger$Department of Physics, Sam Houston State University, Huntsville TX 77341}

\date{\today}

\begin{abstract}
The entanglement entropy of two gapless non-interacting fermion subsystems is computed approximately in a way that avoids the introduction of replicas and a geometric interpretation of the reduced density matrix. We exploit the similarity between the Schmidt basis wavefunction and superfluid BCS wavefunction and compute the entropy using the BCS approximation. Within this analogy, the Cooper pairs are particle-hole pairs straddling the boundary and the effective interaction between them is induced by the projection of  the Hilbert space onto the incomplete Schmidt basis. The resulting singular interaction may be thought of as "lifting" the degeneracy of the single particle distribution function.  For two coupled fermion systems of linear size $L$, we solve the BCS gap equation approximately to find the entropy $S \approx (w^2/t^2)\log{L}$ where $w$ is the hopping amplitude at the boundary of the subsystem and $2t$ is the bandwidth. We further interpret this result based upon the relationship between entanglement spectrum, entropy and number fluctuations.

\end{abstract}

\maketitle
\section{Introduction} 

Quantum entanglement seems to provide an important connection between several distinct fields of physics ranging from conformal field theory \cite{cardy_review2}  and topologically ordered phases in quantum field theories \cite{kitaevpreskill} to quantum gravity  \cite{ryu_review}. One important quantity is the {\sl entanglement entropy}, computed for a finite subregion of a quantum field theory (QFT) or many body system. If, in a QFT in $d$ spatial dimensions, a distinguished region $A$ of volume $L^{d}$ is formed,  it follows that the degrees of freedom which reside exclusively in the region $A$ will appear to be in a mixed state.  The degree of mixing may be characterized by the entanglement entropy, $S = -{\rm tr}\rho \ln{\rho}$, where  the reduced density matrix $\rho = {\rm tr}_{\notin A}|0\>\<0|$ has been formed by tracing over the degrees of freedom of the ground state, $|0\>$, exterior to the region $A$. 

Entanglement entropy typically obeys an {\sl area law} and is proportional to the area of the bounding surface ($L^{d-1}$), although several variants are possible depending on the underlying particle statistics (fermion or boson) and dimensionality.  $1 + 1$-dimensional CFTs---which describe critical spin chains, Luttinger liquids and other massless theories---have pointlike  bounding surfaces;  however, the entanglement entropy was shown to depend universally upon the central charge of the theory and to diverge logarithmically with the length of the distinguished region \cite{CallanWilczek,Holzhey}.  Specifically, the entropy for an open chain is given by  a universal expression, $S =  \frac{c}{3} \ln{L/\epsilon}$ where $c$ is the central charge and $\epsilon$ is a spatial cut-off. 

Entanglement related quantities have also become a way of characterizing strongly correlated and topologic phases in condensed matter physics. Entanglement entropy diverges at a critical point and has been used to identify critical phases \cite{QPT}. Topologic entanglement entropy (a subdominant term in the entropy) and the entanglement spectrum (the complete set  of eigenvalues of the reduced density matrix) have been shown carry information about the statistics of the underlying quasiparticles in topologic phases \cite{top_ent}. Recently, the entanglement spectrum of free fermion models has been linked to the spectrum of edge modes in topologic insulators \cite{e_spect_edge}. Lastly, important connections between noise, full-counting statistics and entanglement in free fermion models has emerged \cite{KL_noise}.

The universal expression for entanglement entropy of a CFT has a purely geometric origin. As originally discovered by Callan and Wilczek \cite{CallanWilczek} (based upon work by Cardy and Peschel \cite{CardyPeschel}) entanglement entropy, computed through a kind of replica trick, is equivalent to thermal entropy of a CFT mapped to a conical manifold with a periodic imaginary time coordinate. The appearance of central charge in the expression for the entropy is a consequence of the conformal anomaly and the curvature singularity at the cone apex. The $\log{L}$ entropy dependence may be thought of as an extensive factor arising from the "volume" of a one dimensional gas that is quantized along radial equal time slices. In effect, radial degrees of freedom are quantized in a box of size $\log{L}$.

But there are several exceptions to the "geometry-entropy" connection---that is, 1-$d$ systems exhibiting entropy proportional to $\log{L}$ but lacking any obvious geometrical constructions for the entropy (specifically they lack translation invariance or even expression as a continuum QFT; see the discussion in reference \cite{cardy_review2}, section $3.3$.) The most striking example is found in the work of Refael and Moore \cite{RefaelMoore} who examined entanglement in spin chains at the infinite disorder fixed point. In this limit, the wavefunction is hierarchically organized into products of maximally entangled RVB-like singlets, and the entropy may be computed by enumerating the pairs that straddle the boundary between subsystems. For the disordered $XY$ model (disordered fermions), the entropy is found to be $\frac{c}{3}\ln{2} \ln{L}$.

This raises the question of whether it is possible to {\sl generally} understand the logarithmic behavior for 1-$d$ entropy by enumeration of entangled pairs (of something). In this manuscript we explore building entanglements by weakly coupling two fermion systems, and enumerating the entangled pairs of fermions as a function of coupling.  In particular, for two coupled gapless fermi systems (see figure 1), states within an energy $w$ (the coupling strength) of the fermi surface should be hybridized and contribute to the entropy. This approach suggests expressing the entropy as a perturbation series in $w$. However, the remarkable exact solution by Eisler and Peschel \cite{exact_imp} shows that this is {\sl impossible}---the entropy is not analytic about $w=0$, preventing a simple perturbative expansion. In this manuscript we suggest that the appropriate description at weak coupling is a collective, many-body effect similar to a phase transition, and to compute the entanglement entropy we adopt a procedure analogous to a BCS mean field calculation.

\begin{figure}[ht]
\includegraphics[height=3cm]{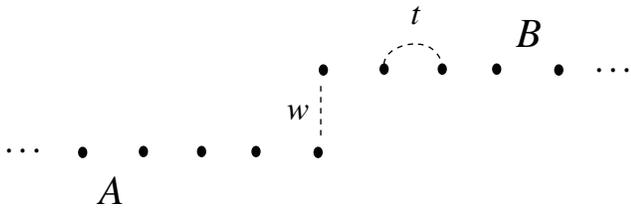}
\caption{ Two coupled fermi chains. Fermions hop between sites with hopping amplitude, $t$, and hop between chains with hopping amplitude, $w$.}
\end{figure}

To understand the origin of the weak coupling singularity, consider two weakly coupled fermi systems as depicted in figure 1. If the fermi systems are noninteracting, Peschel \cite{peschel_corr_fn} has introduced a way of computing the reduced density matrix, $\rho_A$, of subsystem $A$. Following  \cite{peschel_corr_fn}, $\rho_A$ has the form of a free fermion density matrix with a modified lattice kinetic energy
\begin{equation}
\label{rhoA}
\rho_A = \frac{1}{Z}e^{-\sum{K_{xy}c_x^\dagger c_y}}
\end{equation}
where $c_x$ ($c^\dagger_x$) destroys (creates) fermions in subsytem $A$ and obeys the conventional fermion algebra. The eigenvalues of the matrix $K_{xy}$ may, in turn, be computed from the eigenvalues of the ground state free fermion correlation matrix
\begin{equation}
\label{corr}
n_{xy} \equiv \<c_y c^\dagger_x \>
\end{equation}
with $x,y$ restricted to subsystem $A$. Denoting the eigenvalues of $n_{xy}$ by $\{n_k\}$, the eigenvalues of $K_{xy}$ are:
\begin{equation}
\label{dispersion}
K_k = \log{(1/n_k -1)}
\end{equation}

Figure 2 is a depiction of the spectrum of $n_k$ for $w \neq 0$ and $w=0$. The form of $n_k$ is similar to the fermi function at finite temperature ($w \neq 0$) and zero temperature ($w=0$), although the index, $k$, is dimensionless. The entanglement entropy, $S$, is proportional to the width of the distribution $n_k$ at the pseudo fermi surface of one subsystem ($k_F=L/2$ for $L$ fermions in $2L$ total sites) in analogy to thermal entropy at finite temperature. If the subsytems $A$ and $B$ are disconnected, $n_{xy}$ is formed from the unperturbed eigenfunctions of the (disconnected) subsystem $A$:
\begin{equation}
n_{xy}=\sum_{m=1}^{L/2}{\phi_m^A(x)\phi_m^A(y)}
\end{equation}
The eigenfunctions of $n_{xy}$ are simply the complete set of unperturbed free particle eigenfunctions $\{\phi^A_m(x)\}$ with eigenvalues $0$ (for $m>L/2$) or $1$ (for $m<L/2$). Consequently, the eigenvalues of $K_{xy}$ are all singular, leading to the non-analytic behavior of the density matrix (\ref{rhoA}).  

The key to computing the entropy is then to describe how the extensive $O(L)$ degeneracy in the eigenvalues of $n_{xy}$---in effect, the entanglement spectrum---is lifted for nonzero $w$.  We will argue that the degeneracy is resolved in a manner analogous to the weak coupling BCS superconducting instability. Although the original fermions are noninteracting, the projective constraint of the Schmidt basis introduces an effective interaction between particles and holes belonging to different subsystems. Since this pair interaction is singular in energy, a solution to the BCS gap equation results in a gap with a logarithmic behavior: $\Delta \sim O(\log{L}/L)$.
\begin{figure}[ht]
\includegraphics[width=7.5cm]{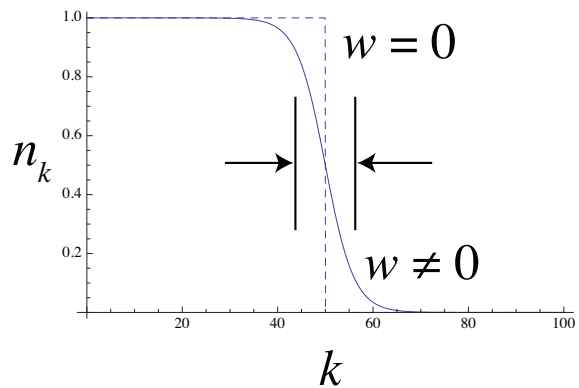}
\caption{\label{fig2} Depiction of eigenvalues of correlation function matrix (eq. \ref{corr}) for an $L =100$ lattice. For $w=0$, the distribution is a step function (dotted line). For $w \ne 0$, the width of the distribution (shown greatly exaggerated) is proportional to the entanglement entropy.}
\end{figure}

The goal of this manuscript is to compute the $\log{L}$ entropy common to many 1-d systems without resorting to replicas and geometric interpretation of the partition function. To accomplish this we exploit the similarity between the Schmidt wavefunction and superfluid BCS wavefunction, where the "glue" that stabilizes the BCS wavefunction is a type of two particle forward scattering across the weak link. The weak link hamiltonian is introduced to organize perturbation theory leading to an effective hamiltonian, which is then solved within a mean field BCS ansatz. Although our solution accomplishes the stated goal---exhibiting the $\log{L}$ entropy---it fails in another way, when compared with exact and numerical results for entanglement entropy.  The entropy of fermionic systems with a weak link has been studied in many ways \cite{peschel_imp1,peschel_imp2,levine_miller,levine_imp}, culminating in the tour-de-force exact solution by Eisler and Peschel  \cite{exact_imp} for noninteracting fermion chains with arbitrary coupling strengths.  For noninteracting chains, the exact solution shows that the entropy remains logarithmic in $L$ even for weakest couplings explored. Early numerical work suggested that entropy was approximately quadratic in the weak link strength---$O(w^2)$ for $w<<t$---but the exact solution exhibited a surprising non-analytic feature, 
\begin{equation}
\label{exact_s}
S \approx (w^2/t^2)\log{(t/w)}\log{L}
\end{equation}
in this limit. In contrast, our mean field solution, which relies on imposing a uniform (momentum independent) gap, yields an entropy, $S \approx (w^2/t^2)\log{L}$. In our conclusion, we speculate on the cause of this discrepancy.

\section{The Schmidt basis for fermions}

In this manuscript, we study the entanglement entropy of a 1-d noninteracting fermion system ($A$) connected by a weak link to a second identical system ($B$). The model hamiltonian is:
\begin{eqnarray}
\label{ham}
H &=& -t\sum_{\< i,j \>; \alpha=A,B}{(c^{\alpha\dagger}_{i}c^\alpha_{j} + c^{\alpha\dagger}_{j}c^\alpha_{i}})\\
\nonumber&-& w (c^{A\dagger}_{1}c^B_{1} + c^{B\dagger}_{1}c^A_{1})
\end{eqnarray}
where $i$ and $j$ are site indices and $A$ and $B$ denote the two identical systems coupled through a hopping amplitude, $w$.  Each subsystem consists of $L$ sites and is taken to have fixed boundary conditions at its ends. Furthermore, we will restrict our calculations to the case of $L$ fermions in $2L$ sites. This 1-d model may also be thought of as the effective 1-d model obtained by coupling two higher dimensional models by a weak link and identifying left/right moving modes with in/out $s$-wave scattering states \cite{levine_miller,swingle}. 
The Schmidt decomposition theorem guarantees that the groundstate of the hamiltonian (\ref{ham}) can be written as a sum with the form:
\begin{equation}
\label{schmidt}
|\psi\rangle = \sum_\alpha{a_\alpha |\alpha\rangle_A  |\alpha\rangle_B}
\end{equation}
where $|\alpha\rangle_{A}$, for instance, are specially chosen states (the Schmidt basis) consisting of linear combinations of states drawn exclusively from subsystem $A$. 

It has been known for some time that a system of free fields (bosonic fields or fermionic fields) partitioned into two nonintersecting spatial subsystems has a coherent state wavefunction that resembles the BCS wavefunction of superconductivity.  This wavefunction dates back to earliest explorations quantum fields in curved space and black hole quantum mechanics by Fulling, Parker, Unruh and Hawking \cite{BirrellDavies}.  Klich and others have independently developed this type of wavefunction in connection with entanglement entropy and noise in condensed matter systems \cite{klich_schmidt,refael_numberfluct}, and the density matrix renomalization group \cite{dmrg_schmidt}.

First consider the hamiltonian (\ref{ham}) with $w=t$, a single, contiguous, tight binding fermion system with length $2L$. We will denote by $\phi_m(x)$ the eigenfunction of the correlation function, $n_{xy}$, with $x$ and $y$ restricted to subsystem $A$ and eigenvalue $n_m$:
\begin{equation}
\label{corr_eqn}
\sum_{y=1}^L{n_{xy}\phi_m(y)} = n_m \phi_m(x)
\end{equation}

Klich \cite{klich_schmidt} has shown that the groundstate wavefunction may be written in the Schmidt form:
\begin{equation}
\label{schmidt} 
|\psi\rangle = \prod_{m=1}^L(\sqrt{1-n_m} + \sqrt{n_m} f_m^{A \dagger}f_{L-m}^B)|0\rangle_A |\bar{0}\rangle_B
\end{equation}
where the fermion operator, $f_m^{A(B) \dagger}$, creates a fermion in mode $\phi_m(x)$ in subsystem $A$ ($B$); the vacuum states $|0\rangle_A$ and $ |\bar{0}\rangle_B$ are the zero particle state of subsystem $A$ and the completely filled state of subsystem $B$, respectively.  Thus the groundstate may be thought of as a BCS condensate of particle hole pairs, but with free particle wavefunctions that are the eigenstates of $n_{xy}$. From this wavefunction, it can shown that the entanglement entropy is proportional to the width of the distribution $n_m$.

Based on the similarity between equation (\ref{schmidt}) and the BCS wavefunction, we attempt to {\sl compute} the entanglement entropy from an appropriate BCS, mean-field starting point.   For a 1-d fermion chain, this would amount to showing that the distribution, $n_m$, akin to the superconductivity coherence factor, $|v_k|^2$, has a width that depends logarithmically on the system size, $L$. 

There is, of course, a significant difference between the interacting fermions in the BCS superfluid and the system of free fermions described by the hamiltonian (\ref{ham}). Attractive interactions between fermions in the superfluid destabilize the fermi sea and lead to the BCS coherent state. Before turning to the question of the pairing "glue" that stabilizes this wavefunction, we mention  a peculiarity of the Schmidt basis that is suggestive of an effective interaction between pairs. Since the correlation function, $n_{xy}$, depends implicitly upon the total number of fermions, $N$, the Schmidt basis is a representation of free fermions only for a specific $N$. Consider the hamiltonian (\ref{ham}) transformed to the Schmidt basis through:
\begin{equation}  
\label{unruh}
f^{A\dagger}_l = \sum_{x \in A}{\phi_l(x) c^{A\dagger}_x} 
\end{equation}
The free particle wavefunctions, $\phi_l(x)$, and eigenvalues, $n_l$, also depend implicitly upon $N$; it follows that the kinetic energy matrix of the transformed hamiltonian depends upon $N$. Within this representation, adding or removing a fermion shifts the energies of all of the other fermions---an effect we typically ascribe to interactions between fermions.

\section{Projection onto Schmidt basis and the pairing interaction}

In this manuscript, we argue that an effective particle-hole pairing interaction stabilizing the wavefunction (\ref{schmidt}) arises from projective constraints imposed upon the free fermion hilbert space. There are many examples of effective interactions arising this way, leading to some of the most interesting phenomena in condensed matter physics: The fractional quantum Hall effect, resonating valence bond antiferromagnetism and superexchange. In the latter case, strong coulomb repulsion removes from the hilbert space states in which two fermions (of opposite spin) occupy the same atomic orbital. The effective low energy theory is the isotropic Heisenberg interaction between the spin degrees of freedom.  In some sense, free particle kinetic energy and the projective constraint of no double occupancy have been "traded in" for an interaction between particles. 

Our central result is that the logarithmically divergent entanglement entropy of the 1-d fermion system (\ref{ham}) may be derived from a weak coupling limit $(w \approx 0)$ by introducing a constraint on the hilbert space of the combined $A$ and $B$ subsystems; namely, it is forced to conform to the many-body Schmidt basis.  The importance of the Schmidt basis for entanglement is that each noninteracting many body fermion state for subsystem $A$ appearing in the wavefunction (a product of $f^{A\dagger}$'s) is correlated with exactly one unique complementary fermion state for subsystem $B$ (a product of $f^{B\dagger}$'s). 

\begin{figure}[ht]
\includegraphics[width=7.5cm]{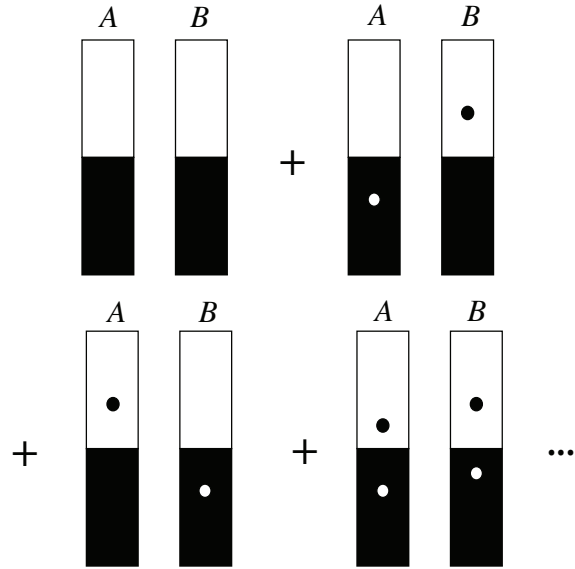}
\caption{\label{allowed} States allowed in the Schmidt basis for $w \neq 0$. The presence of the first state ($N^A=N^B=0$) forces $N^A-N^B$ to be even for all other states contributing to the groundstate, where $N^\alpha$ is the number of particle-hole pairs in system $\alpha$ $(=A,B)$}
\end{figure}

\begin{figure}[ht]
\includegraphics[width=5.5cm]{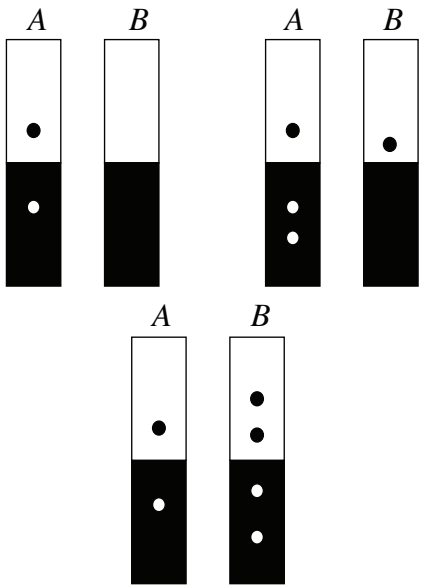}
\caption{\label{disallowed} States not allowed in Schmidt basis. The first state is not allowed because $N^B=0$ already appears uniquely correlated with $N^A=0$ in figure \ref{allowed}. Similarly, the second state is not allowed because the state depicted for $B$ (also an $N^B=0$ state) already appears correlated with a different state of $A$.}
\end{figure}

When the two systems are decoupled ($w=0$), the Schmidt state is trivially:
\begin{equation}
\label{decoupled}
\prod_{k=1}^{L/2}{c_k^{A\dagger}c_{k}^{B\dagger}}|0\rangle_A |0\rangle_B
\end{equation}
where 
\begin{equation}
\label{unperturbed}
c^{\alpha\dagger}_k = \sqrt{\frac{2}{L+1}} \sum_{x=1}^L{\sin{\frac{k\pi x}{L+1}} c^{\alpha\dagger}_x} \end{equation}
(Note that the condition $f^\alpha_k = c^\alpha_k$ may be chosen when the systems are decoupled.) Equation \ref{decoupled} corresponds to the first state depicted in figure \ref{allowed}. Since the state (\ref{decoupled}) continues to contribute to ground state when $w \neq 0$, the next contributions must be states of the type depicted in the second and third states of figure \ref{allowed}. For instance, the first state depicted in figure \ref{disallowed} cannot contribute to the ground state because the zero particle-hole state for system $B$ already appears in the ground state, uniquely correlated with the zero particle-hole state for system $A$. Similarly, the second state in figure \ref{allowed} precludes the second state in figure \ref{disallowed}, and so on.

Denoting the number of particle-hole pairs in system $\alpha$ $(=A,B)$ by $N^\alpha$, the projective constraint on the Hilbert space is
\begin{equation}
\label{constraint}
 N^A - N^B = {\rm even }\,\,{\rm integer} 
 \end{equation}
 Of course, the total number of particles must also be conserved.  

\begin{figure}[ht]
\includegraphics[width=2.5cm]{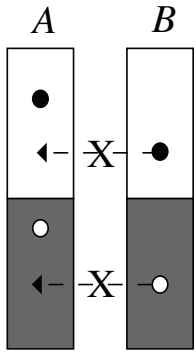}
\caption{\label{blocked} The constraint, $N^A - N^B$ even, may be implemented kinematically by blocking the scattering of positive (negative) energy states in $A$ into positive (negative) energy states in $B$, and vice versa.}
\end{figure}

This constraint is most naturally expressed kinematically. Figure \ref{blocked} depicts two possible processes allowed by the weak link term of the hamiltonian (see eqn. (\ref{wl_momentum}) below): the scattering of a positive energy particle in $A$ into a positive energy state in $B$, and the scattering of a negative energy hole in $A$ into negative energy state in $B$. If these two processes (and their time reversed counterparts) are blocked, scattering occurs only within the constrained subspace. Next we consider how to make an effective hamiltonian that is consistent with the constrained hilbert space.

Following a technique introduced by Anderson for the renormalization group in the Kondo problem \cite{poorman}, an interaction between particles is introduced that  preserves the Dyson equation for the many-body T-matrix in the projected hilbert space.

The details of the projective RG are given in \cite{poorman}; here we aim to give only a brief summary as it applies to the present case (which does not involve RG.) The $T$-matrix, $T(\omega)$, is one way of linking the bare green's function $G_0(\omega)$ to the exact one, $G(\omega)$, and is defined by the following equation:
\begin{equation}
G(\omega) = G_0(\omega) + G_0(\omega)T(\omega) G_0(\omega)
\end{equation}
Here, the green's function is taken to be the full, many-body,  resolvent operator: $G^{-1}=\omega - H$. The $T$-matrix satisfies a Dyson equation:
\begin{equation}
T(\omega) = H_w + H_w G_0(\omega) T(\omega)
\end{equation}
Following \cite{poorman} we try to find a new interaction operator, $V$, replacing the weak link hamiltonian, $H_w$, so that the Dyson equation is satisfied in a projected hilbert space satisfying the constraint (\ref{constraint}). Consider the projection operator, $P$, which projects a state onto the subspace satisfying (\ref{constraint}), and with a fixed total number of particles equal to  $L$. Denote by $\bar{O}$ an operator $O$ that has been projected onto this subspace; that is, $\bar{O} = POP$.  It is shown in \cite{poorman} that the $T$-matrix restricted to the projected subspace satisfies a Dyson equation also restricted to the subspace,
\begin{equation}
\bar{T}(\omega) = \bar{V} + \bar{V} G_0(\omega) \bar{T}(\omega)
\end{equation}
where the effective interaction, to $O(w^2)$, is:
\begin{equation}
\label{effint}
\bar{V} = PH_wP + PH_w(1-P)G_0H_wP + \ldots
\end{equation}

The first term of equation (\ref{effint}) describes the creation of a particle hole pair that straddles the $A/B$ subsystems. The second term of (\ref{effint}) describes fluctuations out of the projected subspace. Singular behavior in energy of the second (fluctuation) term in the effective interaction $\bar{V}$ is the feature that ultimately leads to the logarithmic dependence of the entanglement entropy on $L$.  
\begin{figure}[ht]
\includegraphics[width=5.0cm]{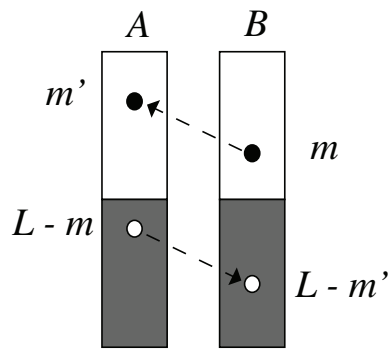}
\caption{\label{scattering} $A$-$B$ particle-hole pair at $(L-m,m)$ scatters into a state $(m^\prime, L-m^\prime)$. The intermediate state consists of a particle hole pair, $(m^\prime, L-m)$, in $A$, existing out of the projected subspace.}
\end{figure}

To analyze this effective interaction, we rewrite $H_w$ in momentum space:
\begin{equation}
\label{wl_momentum}
H_w = -w \frac{2}{L+1}\sum_{m,m^\prime=1}^L{M_{m,m^\prime}(c_m^{A\dagger} c_{m^\prime}^B + c_m^{B\dagger} c_{m^\prime}^A)}
\end{equation}
where 
\begin{equation}
c^{A,B}(x) = \sqrt{\frac{2}{L+1}}\sum_{m=1}^L{\sin{\frac{m\pi x}{L+1}}c_m^{A,B}}
\end{equation}
and $M_{m,m^\prime} = \sin{\frac{m \pi}{L+1}}\sin{\frac{m^\prime \pi}{L+1}}$. The kinetic energy, $H_0$, is:
\begin{equation}
H_0 = \sum_{m=1}^L{\epsilon_m (c_m^{A \dagger}c_m^A + c_m^{B \dagger}c_m^B)}
\end{equation}
where $\epsilon_m = -2t \cos{\frac{m \pi}{L+1}}$.

We consider a situation, as depicted in figure (\ref{scattering}), where an $A$-$B$ particle-hole pair at $(L-m,m)$ scatters into a state $(m^\prime, L-m^\prime)$. The intermediate state consists of a particle hole pair, $(m^\prime, L-m)$, in $A$, existing out of the projected subspace.  Restricting the scattering as described, the relevant term in the effective interaction (\ref{effint}) is:
\begin{eqnarray}
\label{scatt}
\nonumber \tilde{V} &=& \frac{4 w^2}{(L+1)^2}\sum_{m,m^\prime>0}M_{m, m^\prime}^2 \frac{c_{L-m}^{A\dagger} c_{L-m^\prime}^B  c_{m^\prime}^{A\dagger} c_{m}^B}{\omega - (\epsilon_m +\epsilon_{m^\prime})}\\
&=&  \sum_{m,m^\prime}{V_{m,m^\prime} c_{L-m}^{A\dagger} c_{L-m^\prime}^B  c_{m^\prime}^{A\dagger} c_{m}^B}
\end{eqnarray}
This interaction, as in the phonon mediated BCS-theory, is a frequency dependent retarded interaction, and we concentrate on the low frequency behavior  $\omega \approx 0$ where the interaction is attractive. 

\section{BCS Solution of the effective hamiltonian}

When the two subsystems are disconnected ($w=0$), the wavefunction (\ref{schmidt}) degenerates to two disconnected fermi seas as expected. The eigenfunctions of $n_{xy}$, as described above in (\ref{corr_eqn}), become the free particle eigenfunctions of the half-chain, and $n_m = \theta(L/2-m)$.  The wavefunction (\ref{schmidt}) then becomes
\begin{equation} 
\label{weak}
|\psi\rangle = \prod(\sqrt{1-n_m} + \sqrt{n_m} c_m^{A \dagger}c_{L-m}^B)|0\rangle_A |\bar{0}\rangle_B
\end{equation}
where $c_m, c_m^\dagger$ create and destroy simple band fermions in the disconnected subsystems. For weak coupling, $w \approx 0$, the pairing approximation of BCS theory assumes that only the coherence factors $\sqrt{1-n_m}$ and $\sqrt{n_m}$ in (\ref{weak})---these are the familiar $u_k$ and $v_k$ coefficients of superconductivity---are modified from their discontinuous distributions at $w=0$, and the wavefunctions corresponding to $c_m, c_m^\dagger$ are not. 

Following the BCS procedure, we take the wavefunction (\ref{weak}) as a variational wavefunction for the hamiltonian, $H = H_0 + \tilde{V}$. To simplify the notation, constants and the singular part of the effective interaction have been grouped into the scattering amplitude $V_{m,m^\prime}$ on the second line of (\ref{scatt}).  Defining a "gap" function,
\begin{equation}
\label{gap}
\Delta_m = -\sum_{m^\prime} V_{m,m^\prime} \sqrt{n_{m^\prime}(1-n_{m^\prime})},
\end{equation}
the variational energy becomes:
\begin{equation}
\label{variational}
U = \sum_m{(2\epsilon_m n_m - \Delta_m \sqrt{n_m(1-n_m)})} 
\end{equation}
Minimizing the energy, $U$, with respect to $n_m$ and solving for $n_m$ gives coherence factors that take the usual BCS form:
\begin{equation}
\label{coherence_factor}
n_m = \frac{1}{2}(1-\frac{2\epsilon_m}{E_m})
\end{equation}
where $E_m^2 = \Delta_m^2 + (2\epsilon_m)^2$. 

Next we turn to solving the gap equation (\ref{gap}). Although the scattering amplitude, $V_{m,m^\prime}$, is an energy dependent interaction, we will seek an approximate uniform solution to the gap equation.  $V_{m,m^\prime}$ has a maximum value when $m, m^\prime \approx L/2$; that is, when scattering events are close to the fermi points of subsystems $A$ and $B$. We evaluate the scattering amplitude in the static limit, $\omega \approx 0$, and for a low energy initial pair state ($\epsilon_m \approx 0$). Linearizing about the fermi point, $m=L/2$, the kinetic energy is
\begin{equation}
\epsilon_p \approx \frac{2 \pi t}{L} p \,\,\,\,\,\,\,\,\,\, p = \pm1, \pm 2 \ldots
\end{equation}
and the effective interaction, $V_{m,m^\prime}$, becomes:
\begin{equation}
V_{m,m^\prime} \approx -\frac{w^2}{Lt}\frac{1}{m^\prime}
\end{equation}
where $L+1$ has been replaced by $L$ and a numerical factor has been absorbed into $w$.

Following the standard BCS procedure and choosing a uniform solution for the gap, $\Delta = \Delta_m$,
the gap equation (\ref{gap}) becomes:
\begin{equation}
\label{gap2}
1 = \frac{1}{2} \frac{w^2}{Lt}\sum_{m}{\frac{1}{m} \frac{1}{E_m}} 
\end{equation}
To solve the gap equation, we convert the sum (\ref{gap2}) to an integral and cut off the energy integration at a scale, $\alpha t$, where $\alpha (< 1)$ is a numerical constant, giving:
\begin{equation}
\label{gap3}
1 = \frac{1}{4} \frac{w^2}{Lt} \int_{t/L}^{\alpha t}{\frac{d\epsilon}{\epsilon} \frac{1}{\sqrt{\epsilon^2 + (\Delta/2)^2}}}
\end{equation}
Looking for a self-consistent solution, we approximate the integrand by  $2/(\Delta \epsilon)$ and integrate to get:
\begin{equation}
\label{gap_solution}
\Delta \approx \frac{1}{2} \frac{w^2}{Lt} \ln{L} + c(\alpha)
\end{equation}
where $c(\alpha)$ is a constant depending upon the cut-off. This solution is self-consistent in that $\Delta$ is comparable to the energy cut-off (justifying truncation of the integrand) for large $L$. 

Within the Sommerfeld approximation, the entropy is proportional to the product of the gap and the density of states at the fermi point, $N_0=L/2\pi t$, and we arrive at:
\begin{equation}
\label{ent_ent}
S \approx \Delta N_0 \approx \frac{w^2}{t^2}\ln{L}
\end{equation}
This result agrees qualitatively with numerical computations of the entanglement entropy for 1-d fermions in a weak link geometry \cite{peschel_imp1}. Specifically, the entropy is logarithmic and the prefactor is {\sl approximately} quadratic in $w$. Other numerical computations of entropy for 2-d and 3-d systems with weak links (i.e. quasi-one dimensional systems) have also exhibited approximately quadratic behavior \cite{levine_miller}. However, the exact solution (and earlier work) informs us that the prefactor is non-analytic and approximately proportional to $w^2 \ln{(1/w)}$ \cite{exact_imp}. Even in the original numerical work \cite{peschel_imp1}, Peschel noted a very small, persistent deviation from quadratic dependence upon the prefactor.  We return to this point in the Discussion section.

\section{number fluctuations and entropy}

In a 1-d translation invariant fermion system, it has been rigorously established that the entanglement entropy of a subsystem is proportional to the number fluctuations in that subsystem \cite{KL_noise, refael_numberfluct,klich_schmidt,lehur}. For such a continuum 1-d system, the entanglement entropy of a length $L$ domain is proportional to $\log{L}$ with a universal prefactor that depends upon the central charge of the underlying CFT.  For subsystems separated by a bond or site defect, the entropy remains logarithmic but with a modified pre-factor \cite{exact_imp}.  

However, in such systems lacking translation invariance the relationship between entropy and  number fluctuations---more generally, the complete set of cumulants---becomes complicated.  Klich and Levitov \cite{KL_noise} studied entanglement entropy in a gated quantum point contact (QPC) and have shown that entropy may be expressed as an asymptotic series in the cumulants (see also reference \cite{KL_refinement} for refinements.) For a QPC with perfect transmission, all but the first and second cumulants vanish, and entropy is simply proportional to number fluctuations. For barriers with nonzero reflection coefficients, relevant to the weak link geometry, the FCS generating function is non-Gaussian and all even cumulants generally contribute to the entropy. 

In this section we explore the number fluctuation-entropy connection based upon perturbation theory applied to weakly coupled fermion chains. The number fluctuations in one subsystem $\delta N^2 \equiv \langle N^2\rangle - \langle N\rangle^2$ is proportional to $\log{L}$, but with a prefactor that is analytic as $w \rightarrow 0$ (a result derived earlier in \cite{eisler_garmon}). Since the entropy in this limit is known through the exact solution to depend nonanalytically on $w$, number fluctuations and entropy cannot be proportional. The ratio of entropy to number fluctuations turns out to be
\begin{equation}
\label{ratio}
\frac{S}{\delta N^2} \propto \ln{\frac{t}{w}}
\end{equation}
and is divergent as $w \rightarrow 0$.  This result suggests that the nonanalytic behavior of the entropy can only be captured by resumming the logarithmic terms in the cumulant series for the entropy to all orders in $w$.

Fluctuations in $N$ may be computed perturbatively in $w$ using straightforward canonical methods. Defining the density-density correlation function
\begin{equation}
\Pi(x,y,t) = \langle T\rho(x,t)\rho(y,0)\rangle
\end{equation}
where $\rho(x,t) \equiv c^{A\dagger}(x,t)c^{A}(x,t)$, the number fluctuations may be computed from
\begin{equation}
\langle N^2\rangle =\sum_{x,y=1}^L \Pi(x,y,0) 
\end{equation}
Denoting by $G$ the single particle propagator for system $A$ (not to be confused with the many-body resolvent operator $G(\omega)$), $G$ may be written in the interaction picture,
\begin{equation}
G(x,y,t) = -i\langle Tc^A(x,t)c^{A\dagger}(y,0)e^{-i\int_{-\infty}^\infty{H_w(t^\prime)dt^\prime}}\rangle
\end{equation}
where $T$ is the time-ordering symbol and $\langle \ldots \rangle$ refers to average taken in the ground states of the disconnected subsystems $A$ and $B$.  Expanding the propagator to $O(w^2)$,
\begin{eqnarray}
\nonumber G_2(x,y,0^-) &=& -\frac{w^2}{2}\int_{-\infty}^\infty{dt_1} \int_{-\infty}^\infty{dt_2} G_0(x,1,-t_1)\\ &\times& G_0(1,y,t_2)G_0(1,1,t_1-t_2)
\end{eqnarray}
where $G_0$ is the unperturbed propagator
\begin{equation}
G_0(x,y,t) = -i\langle Tc^A(x,t)c^{A\dagger}(y,0)\rangle
\end{equation}
Inserting the expansion of $G$ up to second order into $\Pi$, we obtain
\begin{equation}
\Pi(x,y) = G_0(x,y)G_0(x,y) + G_0(x,y)G_2(x,y) + \ldots
\end{equation}
where we have dropped the time argument, $t=0$. The covariance to $O(w^2)$ is then:
\begin{equation}
\label{dpt}
\delta N^2 \equiv \langle N^2\rangle - \langle N\rangle^2 =\sum_{x,y} G_0(x,y)G_2(x,y) 
\end{equation}
Expressing $G_2$ in spectral form, the covariance may be written:
\begin{equation}
\label{covariance}
\delta N^2 = 2(\frac{w}{L+1})^2 \sum_{m=1}^{L/2}\sum_{m^\prime=1}^L{\delta_{m^\prime}\frac{M_{m,m^\prime}^2}{(\epsilon_m - \epsilon_{m^\prime})^2}}
\end{equation}
where $\delta_m \equiv \rm{sgn}(\epsilon_m)$ and the resonant terms $m=m^\prime$ are excluded. The sum (\ref{covariance}) may be turned into an integral and evaluated to give
\begin{equation}
\label{number_fluct}
\delta N^2 = C \frac{w^2}{t^2}\ln{L}
\end{equation}
Note that this result is qualitatively the same as the logarithmic behavior found for the number fluctuations in an $L$ site subsystem of a translationally invariant chain. Using (\ref{number_fluct}) and (\ref{exact_s}) to compute the ratio of entropy to number fluctuation establishes the result (\ref{ratio}).

We note that fluctuations in the number of fermions, $N$, in subsystem $A$ may also be computed from the Schmidt wavefunction (\ref{schmidt}) within our projective approach. Transforming
\begin{equation}
\langle N^2\rangle = \sum_{x,y \in A} \langle \psi |c_x^{A\dagger} c_x^A  c_y^{A\dagger} c_y^A |\psi \rangle
\end{equation}
to the Schmidt basis using the canonical transformation (\ref{unruh}), the fermion number covariance is found to be:
\begin{equation}
\langle N^2\rangle - \langle  N\rangle^2 = \sum_{m=1}^L{n_m(1-n_m)} \approx \Delta N_0
\end{equation}
Again, within the Sommerfeld approximation, the covariance is proportional to the width of the distribution $n_m(1-n_m)$ appearing in the sum above, and gives the same result as found for the entanglement entropy in (\ref{ent_ent}). Thus within our approach, entropy and number fluctuations are simply proportional.

Lastly, we return to the perturbative number fluctuation results, equations (\ref{covariance}) and (\ref{number_fluct}) and interpret them in a different way. The projective approach described previously was motivated by the observation that the extensive degeneracy in the eigenvalues of the correlation function (they are all 0 or 1) is lifted for any non-zero $w$.  From the shift in degenerate eigenvalues of the correlation matrix (from 0 or 1), the entropy might be computed. Since $G_2$ is the first nonvanishing correction to the correlation function, the correction to the $m$th eigenvalue of the correlation function, $\delta n_{m}$, is found from degenerate perturbation theory 
\begin{equation}
\delta n_{m} = \sum_{xy}{\phi_m^A(x) G_2(x,y) \phi_m^A(x)}
\end{equation}
Note that $\sum_{m=1}^{L/2}{\delta n_m}$ is the same as the expression for the number fluctuations, (\ref{dpt}), and leads to the same results, equations (\ref{covariance}) and (\ref{number_fluct}). 

Thus the perturbative expression for number fluctuations is really the same calculation as degenerate perturbation theory applied to the extensive degeneracy of the correlation function eigenvalues for disconnected chains. However, inspection of the numerical eigenvalues, $n_m$, shows that the perturbative corrections to $n_m$ are not even qualitatively correct, even though the final expression for entropy is correct. According to perturbation theory, $\delta n_m \propto 1/m$ for large $m$, whereas the behavior for $\delta n_m$ seen numerically---and by the BCS approach described in this manuscript---is a "gapped" behavior, where $n_m$ is given by the expression (\ref{coherence_factor}).

If higher order cumulants $\langle N^4\rangle_c, \langle N^6\rangle_c$ etc. are also analytic functions of $w$ about $w=0$, it seems that the distribution $n_m$ can only be obtained by resumming the cumulant series to all orders in $w$. Presumably each relevant term is logarithmically divergent in $L$.

\section{discussion}

The motivation of this work was finding an alternative explanation for the $\log L$ entanglement entropy---one that does not rely upon replicas and a geometric interpretation of the resulting partition function. In the latter approach, entanglement entropy appears as the fictitious thermal entropy associated with the partition function evaluated in a conical geometry (or, equivalently in terms of corner transfer matrices.) In a conical geometry the $\log L$ behavior may be traced to the log-periodic coordinates that diagonalize the free particle kinetic energy in polar coordinates. Schematically, the entropy is evaluated for a partition function,
\begin{equation}
Z = \int{D\phi e^{-S_\alpha}}
\end{equation}
that follows from the free particle action on a cone with completion angle $\beta$ (the inverse temperature):
\begin{equation}
S_\alpha = \int_0^\beta{d\theta } \int_0^L{dr ((\partial_r\phi)^2 + \frac{1}{r^2}(\partial_\theta \phi)^2)}
\end{equation}
The action is diagonalized by
\begin{equation}
\phi(r,\theta) \sim e^{i m \phi} e^{i k \frac{\ln{r/\epsilon}}{\ln{L/\epsilon}}}
\end{equation}
where $L$ is the radial length and $\epsilon$ is a radial cut-off.

The divergent $\log L$ entropy appears as a consequence of each momentum eigenvalue collapsing inversely proportional to its quantization "length," $\log{L/\epsilon}$.  Just as entropy in a massless quantum fluid at temperature $1/\beta$ is $S \approx L/\beta$---the aspect ratio of length and imaginary time dimensions---entanglement entropy becomes $S \approx \frac{1}{2\pi}\log{L/\epsilon}$.

In our calculation, entanglement entropy is computed from the number of particle hole pairs that straddle the boundary between subsystems. The particle hole pairs have an effective attraction and form a BCS type condensate; the product of the energy gap, $\Delta$, and the density of states at the fermi level, $N_0$,  is a measure of the number of participating pairs and, therefore, the entanglement entropy. Within our approximations, the gap, $\Delta \approx  \frac{w^2}{Lt} \ln{L} $, and the density of states $N_0 \approx L/t$, giving our central result:
\begin{equation}
S  \approx \frac{w^2}{t^2}\ln{L}
\end{equation}

Although our calculation produces the logarithmic entropy in a novel way, it only agrees approximately with the exact results of entanglement entropy in a fermion chain. Specifically, our result fails to produce the correct prefactor in the entropy,  the exact result being $S \approx (w^2/t^2)\log{(t/w)}\log{L}$ in the limit $w<<t$. 

To understand this discrepancy we first return to the exact solution at the translationally invariant point $(w=t)$ and consider the reduced density matrix (\ref{rhoA}) and effective single particle eigenvalues, $K_k$ (equation (\ref{dispersion})) describing subsystem $A$. In the translation invariant limit $w=t$, the single particle eigenvalues $K_k$ appearing in the quasi thermal fermi distribution (adopting the notation of reference \cite{peschel_review}) may be thought of as the quotient of an effective linearized energy, $\omega_k \approx \pi t k/L$ and an effective temperature $T = \pi t L^{-1}\ln{L}$. (That is, $K_k = \omega_k/T$.) These results follow from the asymptotics of Peschel \cite{peschel_asymptotics} and are convincingly demonstrated in the numerical work of Eisler et al (see figure 1 of ref. \cite{eisler}). The entropy is thus the product of the temperature, $T$, and density of states, $L/\pi t$, giving $S \approx \log{L}$.  Our result, although arrived at in a completely different way (and in a completely different limit, $w<<t$), resembles this result by identifying the effective temperature, $T$, with our gap, $\Delta$.  We also note that BCS type distribution, $n_k$, with gap $\Delta$ in our result, and a fermi distribution with effective temperature $T=\Delta/2$ agree to $O(\epsilon_k^2/\Delta^2)$ and thus give the same result for the entropy in the limit $L \rightarrow \infty$. The two distributions yield identical entropies in that that their widths, which determine the entropy, both vanish as $O(L^{-1}\log{L})$. In summary, our result for the distribution $n_k$, derived in weak coupling $w<<t$, resembles the $n_k$ obtained in the exact solution for $w=t$. 

The discrepancy of our result with the exact result at {\sl weak} coupling $(w << t)$ may be traced to a qualitative change in the effective single particle eigenvalues, $K_k$, when $w<<t$.  In the exact solution, $\omega_k$ develops a gap approximately of the form $\omega_k = \sqrt{\tilde{\Delta}^2 + \epsilon_k^2}$, for $w<<t$. The gap $\tilde{\Delta}$ (distinct from our coherence gap, $\Delta$), is approximately $\tilde{\Delta} \approx \ln{(t/w)}$.  This is the origin of the $\ln{(t/w)}$ appearing in the prefactor of the entropy. In the BCS approach, the distribution $n_k$ (comparable to the fermi distribution in the latter analysis), involves only the free particle kinetic energies of the lattice fermions which are, of course, ungapped.  Although this appears to be a fundamental limitation of our calculation, perhaps an improved variational approach might capture the gap feature.

To summarize these relationships: the computation of both number fluctuations and entropy based upon the Schmidt wavefunction amount to evaluating the width of the distribution $n_k$. In the exact solution, when $w =  t$, $n_k$ has the form of a Fermi distribution with a linear single particle dispersion resembling that of free fermions and an effective temperature $T = \pi t L^{-1}\ln{L}$.  Within the projective/BCS type approach described in this manuscript, $n_k$ has a gapped behavior given by equation (\ref{coherence_factor}) with an approximate linear single particle dispersion given by the original lattice fermions and a gap, $\Delta \approx (w^2/t)L^{-1}\ln{L}$. Thus these two results qualitatively agree as $w \rightarrow t$, given the vanishing width of $n_k$ in both cases as $L \rightarrow \infty$.  On the other hand, the computation of number fluctuations (or entropy) from canonical perturbation theory leads to a result proportional to $\log{L}$, although $n_k$ has a qualitatively different behavior---algebraic rather than gapped.  Our projective approach is then an improvement over perturbation theory, and in qualitative agreement with $n_k$ in the translation invariant case, but, as discussed in detail above, fails to agree with the exact solution when $w<<t$. 

One might argue that BCS theory of superfluidity is mean field theory and is only expected to agree approximately, especially in the case of a one dimensional exact result. Specifically, BCS captures a thermodynamic phase transition, where the order parameter (the phase of the wavefunction representing the condensate) becomes fixed, stabilized by the free fluctuations in the conjugate variable, $N$, (the number of particles in a patch of size equal to the coherence length.) The BCS approximation fails badly in one dimension and marginally in two dimensions because phase fluctuations are too strong. Similarly, in a finite size sample, phase fluctuations are not frozen and, in fact, correspond to the zero point motion of a massive rotator.  Since our computation is aimed at finding the finite size scaling of entropy, one might question using the mean field result where fluctuations of the order parameter have been frozen out. Moreover, our BCS approach entails a solution to the energy dependent gap equation that imposes a uniform gap; this also may be responsible for its failure to reproduce the correct nonanalytic singularity.

The main ingredient of our result is the projective constraint on the underlying hilbert space---an approach that might be used in any system where a variational form of the Schmidt basis is postulated. Another advantage to this approach is the focus upon computing $n_k$, in particular finding a reasonable (and physical) approximation for this highly singular distribution.  For noninteracting fermions, $\{n_k\}$ encodes all eigenvalues of the reduced density matrix and therefore the entire entanglement spectrum (a comparable argument exists for bosons). Understanding how the degeneracy in $n_k$ is generically resolved at weak coupling might provide a path to computing entropy in higher dimensions where specialized 1-d methods cannot be used. It is also interesting to consider if the non-analytic feature in the entropy at weak coupling survives in higher dimensions. If fermions interact in such a way that the system is still described as a fermi liquid, this technique might also provide an approach for computing the entanglement spectrum in such systems. The approximation in this case might resemble the Bogolubov-de Gennes \cite{bdg} formalism where interactions are decoupled into off-diagonal Hartree terms to describe superfluidity and diagonal Hartree terms to describe residual interactions.

Lastly, our perturbative calculation of the second cumulant suggests an investigation of the cumulant series for entanglement entropy.  Since the second cumulant is analytic in $w$, it seems necessary to re-sum the series to all orders to capture the nonanalytic behavior exhibited by the exact solution. 

We thank the referees for pointing out the origin of the discrepancy with the exact solution and references \cite{eisler_garmon,eisler}. This research was supported by grants from the Department of Energy DE-FG02-08ER64623---Hofstra University Center for Condensed Matter, Research Corporation CC6535 (GL) and NSF Grant no. 0705048 (BF).

\end{document}